\documentclass[twocolumn,showpacs,amsmath,amssymb,nofootinbib,pra,superscriptaddress]{revtex4-1}

\usepackage[english]{babel}

\usepackage{graphicx} 
\usepackage{xcolor}
\usepackage{mathtools}
\usepackage{amsmath,amssymb}

\begin{document}

\title{Two Cold Atoms in a Time-Dependent Harmonic Trap in One Dimension}

\author{M. \surname{Ebert}}
\affiliation{Institut f{\"u}r Kernphysik, Technische Universit{\"a}t Darmstadt, 64289 Darmstadt, Germany}
\author{A. \surname{Volosniev}}
\affiliation{Institut f{\"u}r Kernphysik, Technische Universit{\"a}t Darmstadt, 64289 Darmstadt, Germany}
\author{H.-W. \surname{Hammer}}
\affiliation{Institut f{\"u}r Kernphysik, Technische Universit{\"a}t Darmstadt, 64289 Darmstadt, Germany}
\affiliation{
ExtreMe Matter Institute EMMI, GSI Helmholtzzentrum f{\"u}r Schwerionenforschung GmbH, 64291 Darmstadt, Germany}

\begin{abstract}
We analyze the dynamics of two atoms with a short-ranged pair interaction in a 
one-dimensional harmonic trap with time-dependent frequency. Our analysis is focused on two representative cases: 
(i) a sudden change of the trapping frequency from one value to another, and (ii) a periodic trapping 
frequency. In case (i), the dynamics of the interacting and the corresponding non-interacting systems turn out to be similar. 
In the second case, however, the interacting system can behave quite differently, especially close to parametric resonance. For instance, in the regions where such resonance occurs we find that the interaction can significantly reduce the rate of energy increase. The implications for applications of 
our findings to cool or heat the system are also discussed.
\end{abstract}

\maketitle

\section{Introduction}

Cold atomic gases offer a unique opportunity to prepare and study few-body systems
in controlled, tunable environments \cite{Serwane2011}. These systems are usually investigated in stationary external traps, 
which motivates the theoretical studies of few-body systems in different stationary confinements 
(see, e.g., Ref.~\cite{zinner2015} for a review of the recent work in one-dimensional (1D) systems). A complementary investigation can be provided by modulating the external traps in time \cite{jin1996}. This possibility drives the studies of time-dependent systems \cite{castin1996,kagan1996,thomas1997,georg2000, Vignolo2011,Schmiedmayer2015}, which are still not understood at the same level of detail as their stationary counterparts.

In this work, we investigate two particles interacting via the delta function potential
in a time-dependent 1D harmonic trap. In the absence of the interaction, the behavior of this system 
is well-understood~\cite{Husimi1953,perelomov1969}. In the presence of the interaction, however, the two-body dynamics 
requires a thorough investigation~\cite{saenz2013,Freyberger2014}. The main objective of our work is to study this dynamics for two representative 
scenarios. In the first scenario, the frequency of the trap changes rapidly from one value to another.  In this case, 
our study suggests that the system behaves pretty much as if it were non-interacting. To explain this behavior, 
we note that the effect of the interaction on the relevant time-independent properties of the system, 
such as energy differences and transition elements, is minute. Therefore, a fast transition from one trapping frequency 
to another should lead to dynamics qualitatively similar to the non-interacting case. In the second scenario, the trapping frequency varies periodically in time. As we discuss in the text, this periodic change can enhance minute changes of the time-independent characteristics of the system. 
Therefore, the interaction can significantly modify the long-time evolution of the system. This scenario bears some similarity to a particular variant of quantum control for the motion of two trapped atoms studied in Ref.~\cite{Freyberger2014}. Another important aspect of this scenario is 
that a periodically driven harmonic oscillator produces unstable parameter regions, i.e., regions in which the energy increases exponentially in time. It is important to understand the role of the interaction in these regions, because, in practical applications parametric resonances can be used to heat or cool the system  (see Refs.~\cite{Modugno2002, Yabuzaki2003} and references therein).  

The paper is organized as follows.
In Sec.~\ref{sec:form},  we formulate the problem and introduce some formalism.  The non-interacting case is addressed
in Sec.~\ref{sec:non_int}. In Sec.~\ref{sec:int_case},  we investigate the interacting case. Finally, we summarize  and give a brief overview of 
possible extensions of this study in Sec.~\ref{sec:concl}. Some technical details are provided in the appendices.

\section{Formulation of the problem}
\label{sec:form}

We consider two particles of masses $m_1$ and $m_2$ in a harmonic trap with time-dependent frequency $\omega(t)$. We always assume that the trapping potential is stationary for~$t<0$ (i.e., $\omega_{t<0} \equiv \omega_0$) and has some time dependence for~$t\geq 0$. Furthermore, we assume that the particles interact via a short-range repulsive
interaction which can be modeled by the potential $V=g\delta(x)$, where $\delta(x)$ is the Dirac delta function
and $g>0$ is the interaction strength.  Note that the assumption of a zero-range interaction is appropriate only for de Broglie 
wavelengths large compared to the range of the potential \cite{olshanii1998}. It can be used to describe low-lying excited states \cite{jochim2012}, 
but is expected to fail for higher excited states. In principle, the description of the low-lying excited states can be improved by adding delta functions
with derivatives but this will not be attempted here.

This two-body system has the Hamiltonian 
\begin{equation}
\mathcal{H}=\sum_{i=1}^2\left(-\frac{\hbar^2}{2m_i}\frac{\partial^2}{\partial q_i^2}+\frac{m_i \omega^2(t) q_i^2}{2}\right)+ V(q_1-q_2) \; ,
\end{equation}
where $q_i$ denotes the position of the particle $i$. To proceed, we decouple the relative motion from the center-of-mass motion by introducing a new set of variables: $x=(q_1-q_2)$ and $R=(m_1 q_1+m_2 q_2)/(m_1+m_2)$. In these variables, the Hamiltonian has the form $\mathcal{H}=H_r+H_{CM}$. Here the $H_r$ describes the relative motion
\begin{equation}
H_r=-\frac{\hbar^2}{2\mu}\frac{\partial^2}{\partial x^2}+\frac{\mu\omega^2(t)x^2}{2}+V(x),
\end{equation}
where $\mu=m_1m_2/(m_1+m_2)$ is the reduced mass. In turn, the $H_{CM}$ describes the center-of-mass motion 
\begin{equation}
H_{CM}=-\frac{\hbar^2}{2 (m_1+m_2)}\frac{\partial^2}{\partial R^2}+\frac{(m_1+m_2)\omega^2(t)R^2}{2}.
\end{equation}
In the following, we only consider $H_r$ because $H_{CM}$ can be treated analogously. 

 In this paper, we assume that for $t<0$ the system occupies a normalized eigenstate of $\mathcal{H}$: $\Phi_r(x)\Phi_{CM}(R)$. 
 This assumption allows us to study the reaction of the eigenstates to the change of the trap. Note also that the ground state of the 
trap has recently been realized experimentally~\cite{jochim2012}. This serves as our motivation to numerically simulate a system which at $t=0$ is in the 
ground state (cf.~Sec.~\ref{sec:int_case}).

The dynamics for $t\geq 0$ is then governed by 
\begin{equation}
i\hbar\frac{\partial \Psi_r(x,t)}{\partial t}=H_r\Psi_r(x,t), \quad \Psi_r(x,t=0)=\Phi_r(x).
\label{eq:schr}
\end{equation} 
In the remainder of the paper, we set $\hbar=\mu=\omega_0=1$ and 
omit the subscript '$r$' for simplicity.

\section{Non-Interacting case}
\label{sec:non_int}

In this section, we focus on the non-interacting system, i.e., $V=0$. This system is interesting because it contains the information about the dynamics induced by the time-dependent trap. This information will be needed later to isolate the effect of the interaction. Moreover, the non-interacting problem can be reduced to the solution
of an ordinary differential equation, see, e.g., Refs.~\cite{Husimi1953, perelomov1969}. Therefore, it gives a reference point for our studies of the
interacting case which require the solution of a partial differential
equation. 

{\it Propagator.}
We start by finding the propagator $K$ of the non-interacting system
\cite{Husimi1953, Khandekar1975}. First note that for the oscillator potential the propagator $K$ is determined by the classical action \cite{feynman_pathint} as $K({\bf2},{\bf1})\sim\mathrm{exp}\left[i\int L \mathrm{d}t\right]$.\footnote{For the oscillator trap, the fluctuations around the classical trajectory in the path integral only contribute to the normalization.} In this expression $L$ is the classical Lagrangian and the integration is taken along the classical trajectory from the point ${\bf 1}$ to ${\bf 2}$. In addition, the proportionality factor can be obtained using the van Vleck-Pauli formula for the quadratic Lagrangian (cf.~Ref.~\cite{JUNKER1985195}). Therefore, all we need to know to find $K$ is $\int L \mathrm{d}t$.
To calculate $\int L \mathrm{d}t$ we note that the trajectories of a particle in a classical time-dependent harmonic oscillator can be mapped onto the ones in a time-independent oscillator, see, e.g., Refs.~\cite{arnoldbook, kanasugi1995, aldaya2011}. Indeed, the classical time-dependent oscillator
\begin{equation}
\frac{\mathrm{d}^2 x}{\mathrm{d}t^2}=-\omega^2(t) x,
\end{equation}
is cast into a stationary oscillator, 
$\frac{\mathrm{d}^2y}{\mathrm{d}\tau^2}=-y$, by going to the scaled variables 
\begin{equation}
\tau=\int_0^t \mathrm{d} k/\lambda^2(k), \qquad y=x/\lambda(t).
\label{eq:scale}
\end{equation}
Here the scale factor $\lambda(t)$ satisfies the equation
\begin{equation}
\lambda^3(t)\frac{\mathrm{d}^2\lambda(t)}{\mathrm{d}t^2}=1-\omega^2(t)\lambda^4(t),
\label{eq:lambda}
\end{equation} 
with the initial conditions $\lambda(0)=1$ and $\dot{\lambda}(0)=0$ ($\dot{\lambda}(t)\equiv \frac{\mathrm{d}\lambda}{\mathrm{d}t}$). Note that the equation for $\lambda$ can be interpreted as Newton's equation for a particle in an inverse square potential and a time-dependent harmonic potential. This observation implies that $\lambda(t)>0, \forall t$. Positivity of $\lambda$ ensures that the transformation to the scaled variables is well-defined. 

Although the very existence of this transformation allows one to understand the classical problem better, in practice, the transformation to the scaled variables does not appear to be very useful because Eq.~(\ref{eq:lambda}) needs to be solved to obtain a solution to the initial problem.  In contrast, we demonstrate below that this transformation allows us to reduce the complexity of the corresponding quantum mechanical problem.

It is straightforward to show that $\int L\mathrm{dt}=\int \tilde L \mathrm{d}\tau + \frac{\dot \lambda x^2}{2 \lambda}|^{\bf 2}_{\bf 1}$, where 
$L =\frac{1}{2}\left(\frac{\mathrm{d}x}{\mathrm{d}t}\right)^2-\frac{1}{2}\omega^2(t)x^2$ and
$\tilde L =\frac{1}{2}\left(\frac{\mathrm{d}y}{\mathrm{d}\tau}\right)^2-\frac{1}{2}y^2$. The classical action for the stationary oscillator
is well-known~\cite{feynman_pathint}, therefore, the scale transformation 
(\ref{eq:scale}) allows us to determine $\int L\mathrm{dt}$ as a function of $\lambda, t$ and ${\bf 1},{\bf 2}$. The integral $\int L\mathrm{dt}$  
in turn determines the propagator  $K(x,t;x',t')$ ($t\geq t'$) for the quantum motion in a time-dependent harmonic trap

\begin{widetext}
\begin{align}
K(x,t;x',t')=\frac{e^{-i\frac{x'^2\dot{\lambda}(t')}{2\lambda(t')}}e^{i\frac{x^2\dot{\lambda}(t)}{2\lambda(t)}} }{\sqrt{\pi\lambda(t')\lambda(t)}} \frac{\text{exp}\left[\frac{i}{2}(\tau(t')-\tau(t))\right]}{\sqrt{1-\text{exp}\left[2i(\tau(t')-\tau(t))\right]}} \text{exp}\left\{i \frac{\left[\left(\frac{x^2}{\lambda^2(t)}+\frac{x'^2}{\lambda^2(t')}\right)\text{cos}\left[\tau(t)-\tau(t')\right]-2\frac{xx'}{\lambda(t)\lambda(t')}\right]}{2\text{sin}\left[\tau(t)-\tau(t')\right]}\right\}.
\label{eq:propagator}
\end{align}
\end{widetext}
Using this propagator we write down an integral equation for the wave function
\begin{equation}
\Psi(x,t)=\int K(x,t;x',0) \Psi(x',0)\mathrm{d}x'.
\end{equation}
By assumption $\Psi(x,0)$ is an eigenstate of the harmonic oscillator, whic
h we denote by $\phi_n$. Therefore, we obtain
\begin{equation}
\Psi(x,t)=\psi_n(x,t)\equiv\frac{e^{-iE_n \tau(t)}}{\sqrt{\lambda(t)}}e^{i\frac{x^2\dot{\lambda}(t)}{2\lambda(t)}}\phi_n\left(\frac{x}{\lambda(t)}\right),
\label{eq:time_ev}
\end{equation}
where $E_n$ is the energy that corresponds to the state~$\phi_n$.
The function $\psi_n$ describes the time evolution of a non-interacting system that initially is in the $n$th state of the harmonic trap. It is interesting to note that, because the set $\{\phi_n\}$ forms a full basis and we have one-to-one correspondence between $\phi_n$ and $\psi_n$, the propagator can be written as $K(x,t;x',0)=\sum \psi_n(x,t)\phi^*_n(x')$ or $K(x,t;x',t')=\sum \psi_n(x,t)\psi^*_n(x',t')$. The equivalence of this form and Eq. (\ref{eq:propagator}) can be checked directly, e.g., using the Mehler formula.

  To complete the discussion of the propagator, we present the time dynamics of a general state. We assume that at $t=0$,
the state is described by the wave function $b(x,t)$ in a stationary trap. In a time-dependent trap, this state then evolves as
\begin{equation}
\Psi(x,t)=\frac{e^{i\frac{x^2\dot{\lambda}(t)}{2\lambda(t)}}}{\sqrt{\lambda(t)}}b\left(\frac{x}{\lambda},\tau\right)\; .
\label{eq:psievol}
\end{equation} 
This result allows us to write down an expression for a Gaussian wave packet 
\begin{equation}
b(x,0)=\frac{1}{\pi^{1/4}}e^{-\frac{(x-x_0)^2}{2}}\;,
\end{equation} 
whose time evolution is given by
\begin{equation}
b(x,t)=\frac{1}{\pi^{1/4}}e^{\frac{-it}{2}}e^{-\frac{1}{2}\left(x^2+\frac{x_0^2(1+\cos2t-i\sin 2t)}{2}-2x_0x(\cos t- i \sin t)\right)} \;.
\end{equation}
Together with Eq.~(\ref{eq:psievol}) this gives immediately $\Psi(x,t)$. 
In the remainder of the paper, however, we focus exclusively on the eigenstates
of the trap. This helps us to isolate the time dynamics due to
the time dependence of the trap. A detailed analysis of the time dynamics of Gaussian wave packets  
is left for future studies (See also Ref.~\cite{Freyberger2014}).

{\it Some properties of the dynamics}. First of all we note that the parity of the wave function is conserved. Thus, in the present manuscript we work only with the subspace $x\geq0$. With this convention, the initial states for the infinitely strong interaction, $1/g=0$,
coincide with the odd parity eigenstates of the non-interacting system~\cite{busch1998}. Moreover, the odd $\psi_n$ vanish at $x=0$; therefore, the time evolution of a system with infinitely strong interactions is also described by Eq.~(\ref{eq:time_ev}). For relevant studies see Refs.~\cite{girardeau2000, castin2004,moroz2012,volosniev2015}.
On the other hand, systems with $0<g<\infty$ have boundary conditions that 
differ from the non-interacting system. Thus, these systems require a different approach which will be discussed in the next section.

\begin{figure}
\centerline{\includegraphics[scale=0.85]{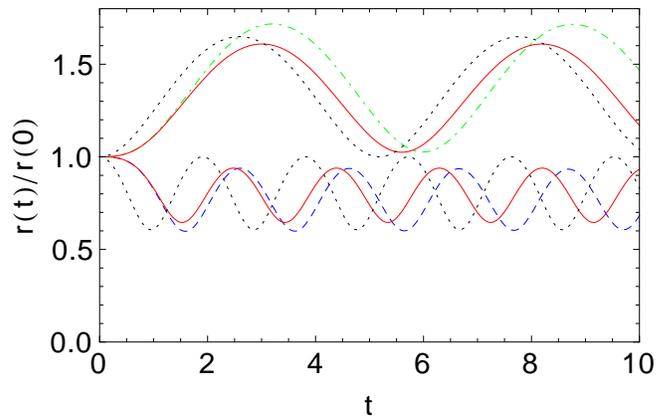}}
\caption{(Color Online). The rescaled root-mean-square radius, $r(t)/r(0)$, as a function of time for $g=1$ and $\alpha=1, T=1$ (blue) dashed, and for $g=1$ and $\alpha=-1, T=1$ (green) dot-dashed curves.  The (red) solid lines show the corresponding non-interacting [infinite interaction] cases, $g=0$  [$1/g=0$]
where $r(t)/r(0)= \lambda(t)$. For comparison, with (black) dotted lines, for the non-interacting cases we show the results obtained using the sudden approximation.  The time-dependence of the frequency is given by Eq. (\ref{eq:omega_exp}).}
\label{fig:re}
\end{figure}

The time dynamics in the non-interacting case is  determined by the function $\lambda(t)$ and $\phi_n$, see Eq.~(\ref{eq:time_ev}). Moreover, for some observables only the knowledge of $\lambda(t)$ is needed. For instance, the root-mean-square radius is written as
\begin{equation}
r(t)\equiv\sqrt{\int x^2 \Psi^*(x,t)\Psi(x,t)\mathrm{d}x}=\lambda(t)r(0).
\end{equation} 
Another interesting quantity is the expectation value of the Hamiltonian,
namely the energy:
\begin{equation}
\epsilon(t)\equiv\int \mathrm{d}x \Psi^* H \Psi = r(0)^2\left(\frac{1}{2}\dot \lambda^2 + \frac{\lambda^2 \omega^2(t)}{2}+ \frac{1}{2 \lambda^2(t)}\right),
\label{eq:epsilon_ni}
\end{equation}
where the expression in the parentheses is the Hamiltonian for the classical problem. By taking the time derivative of Eq.~(\ref{eq:epsilon_ni}), we obtain the rate of the energy change
\begin{equation}
\frac{\mathrm{d}\epsilon}{\mathrm{d}t}=r^2(t)\omega(t)\frac{\mathrm{d}\omega(t)}{\mathrm{d}t}=\frac{r^2(0)}{2}\frac{\mathrm{d}\omega^2(t)}{\mathrm{d}\tau}.
\end{equation} 
Alternatively, this expression can be obtained from the Hellmann-Feynman theorem.

{\it Behavior of $\lambda$ and $\tau$.}
We would like to obtain further insight into the properties $\lambda(t)$ and $\tau$. To simplify the discussion let us note that $\lambda(t)$ can be interpreted as the radial coordinate of a particle in a two-dimensional harmonic oscillator. To show this, we consider the complex function $\xi(t)=\lambda(t)e^{i\tau(t)}$ which obeys the equation
\begin{equation}
\frac{\mathrm{d}^2 \xi}{\mathrm{d} t^2}=-\omega^2(t)\xi(t),
\label{eq:label}
\end{equation}
with $\xi(0)=1$ and $\dot\xi(0)=i$. If we write $\xi=\xi_R+i\xi_I$, then  $\ddot \xi_{R,I}=-\omega^2 \xi_{R,I}$ with $\xi_R(0)=\dot \xi_I(0)=1$ and $\dot \xi_R(0)=\xi_I(0)=0$. Hence, if we consider $\xi_R$ and $\xi_I$ as two independent coordinates we conclude that $\lambda(t)=\sqrt{\xi^2_R+\xi^2_I}$ describes the radial motion of a two-dimensional oscillator; $\tau$ then represents the angular motion. 

\begin{figure}
\centerline{\includegraphics[scale=0.85]{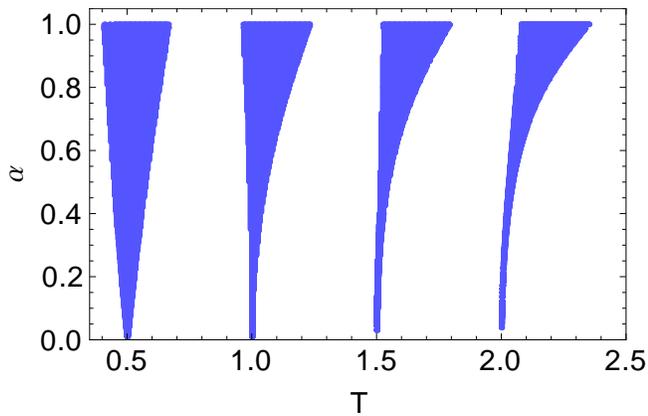}}
\caption{(Color Online). The stability diagram for the case of a periodic change 
of the frequency, i.e.,  $\omega^2(t)=1+\alpha \sin(t/T)$. The solutions of Eq.~(\ref{eq:label}) are 
unstable in the shaded regions.}
\label{fig:lambda2}
\end{figure}

Equation (\ref{eq:label}) is very well studied in the physics and mathematics literature. We present its solutions for two representative cases. The first 
case is an exponential change of the frequency in the beginning and constant frequency later
\begin{equation} \omega^2(t) =
  \begin{cases}
    e^{\alpha t}       & \quad \text{if } 0\leq t\leq T, \\
    e^{\alpha T}  & \quad \text{if } T < t ,\\
  \end{cases}
\label{eq:omega_exp}
\end{equation}
where the parameter $\alpha$ determines the rate of the change. Note that, for $T<t$ Eq.~(\ref{eq:label}) is a time-independent oscillator with the solutions $e^{\pm ie^{\alpha T/2} t}$. Moreover, for $0\leq t\leq T$, Eq.~(\ref{eq:label})
can be transformed to the Bessel equation, with the solution
$\xi(t) = c_1 Y_0\left( \frac{2}{\alpha} e^{\frac{\alpha t}{2}} \right) + c_2 J_0\left(\frac{2}{\alpha}e^{\frac{\alpha t}{2}}\right)$,
where $J_i$ ($Y_i$) is the nth-order Bessel function of the first (second) kind~\cite{abram}. To satisfy the initial conditions, we set $c_1= \frac{\pi}{\alpha}\left[i J_0 \left(\frac{2}{\alpha}\right) + J_1 \left(\frac{2}{a}\right) \right]$ and $c_2= \frac{\pi}{\alpha}\left[-i Y_0\left(\frac{2}{\alpha}\right) - Y_1 \left(\frac{2}{a}\right) \right]$. An example is given in Fig.~\ref{fig:re}, where the solid (red) curves show the behavior of $\lambda(t) = r(t)/r(0)$ for two sets of parameters. 
 For $t>T$ the root-mean-square radius is a periodic, oscillatory function of $t$, because the trap is a stationary oscillator with a period 
proportional to $\exp(-\alpha T/2)$ for these times. For comparision,  we also show the results for a 'true' sudden approximation in Fig.~\ref{fig:re}, i.e., the case where the trap frequency is changed instantly to the final value $\exp(\alpha T)$. We see that the oscillations for the exponential change of frequency are similar to those obtained using a sudden approximation but with a somewhat larger amplitude. The curves for the interacting case are discussed in the next section.

Finally, we study the $\lambda$ and $\tau$ for a periodic oscillator frequency, i.e., $\omega^2(t)=1+\alpha \sin(t/T)$. In this case, Eq.~(\ref{eq:label}) is the Mathieu differential equation with the well-known solution~\cite{abram}.
This solution is very different from the previous one because it assumes the resonant behavior for some parameters\footnote{Throughout this paper, parametric resonance is the phenomenon with $\lim_{t\to\infty}\epsilon(t)\to\infty$.} (cf.~Ref.~\cite{landau_book1, Yakubovich}). In the resonant cases the $\lambda(t)$ is an oscillating function with an amplitude which becomes exponentially large at large times. One way to characterize this behavior is by considering the Floquet solutions~\cite{abram} to Eq.~(\ref{eq:label}), which are of the form $\xi_F(t)=e^{i\gamma t}F(t)$, where $F(t)$ is a periodic function. Obviously, the condition $\mathrm{Im} \gamma<0$ leads to parametric resonance. 
We present the regions where this condition is fulfilled in Fig. \ref{fig:lambda2}. See  Refs.~\cite{abram,Yakubovich} for further discussion.

\begin{figure}
\centerline{\includegraphics[scale=0.85]{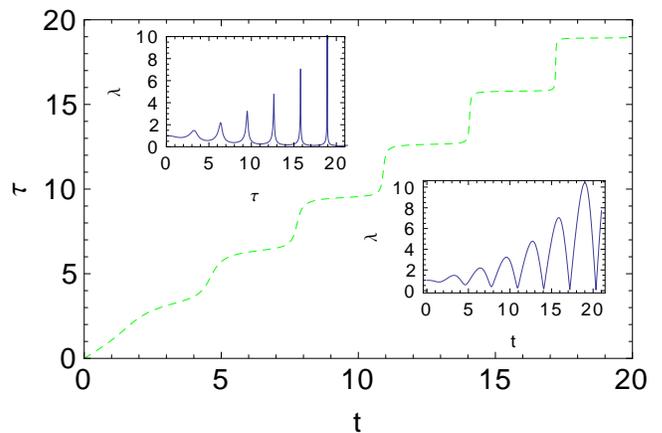}}
\caption{(Color Online). The dependence of $\tau$ on $t$ for $\alpha=0.5$ and $T=0.5$. The insets show the corresponding dependence of $\lambda$ on $t$ and $\tau$.}
\label{fig:tau}
\end{figure}

It is interesting to note that the decomposition $\xi(t)=\lambda(t)e^{i\tau(t)}$ allows us to argue that at resonance for long times, where the divergent Floquet solution dominates, we have  $\tau(t+n \pi)-\tau(t) \sim n \pi$, where $n=0,1,2,...$.  The evolution of $\tau$ with time $t$ at resonance is shown in Fig.~\ref{fig:tau}. 
In the insets, we also plot $\lambda$ as a function of time $t$ and as a function of $\tau$. Clearly, $\tau$ changes significantly only when $\lambda$ is small which leads to a staircase-like behavior. Also note that in the $\tau$-domain the effect of $\lambda$ diminishes, i.e., the region where $\lambda$ is large becomes smaller than in the $t$-domain.

\section{Interacting case}
\label{sec:int_case}

As we have just shown the non-interacting (infinite-interaction) problem 
can be reduced to the solution of an ordinary differential equation.
This reduction is due to the mapping of the time-dependent classical problem onto a simple time-independent problem. Unfortunately, so far it appears impossible to implement a mapping of the interacting problem onto another problem with a known solution. That is why other methods of analysis should be used. In this section, we discuss how to construct a solution as a series. This construction is useful if the excitation of the system is small. Also, we present a numerical routine which is used when perturbative techniques fail.

To construct a series we use the non-interacting propagator in Eq. (\ref{eq:propagator}), which leads to the following integral equation for the $\Psi(x,t)$

\begin{widetext}
\begin{equation}
\Psi(x,t)=\int \mathrm{d}x' K(x,t;x',0)\Psi(x',0)-i\int_0^t \mathrm{d}t'\int \mathrm{d}x' K(x,t;x',t')V(x')\Psi(x',t').
\label{eq:psi}
\end{equation}
\end{widetext}
Iterative solution of this equation gives the desired series expansion.
However, this equation is given in the variables $(x,t)$. As we learned in the previous section it might be advantageous to use the variables $(y,\tau)$ instead. Let us use these variables in Eq.~(\ref{eq:schr}) together with the substitution $\Psi(x,t)=e^{i\frac{x^2\dot{\lambda}(t)}{2\lambda(t)}} f(y,\tau)/\sqrt{\lambda(t)}$. The function $f$ then satisfies the equation 
\begin{equation}
i\frac{\partial f}{\partial \tau}=-\frac{1}{2}\frac{\partial^2}{\partial y^2} f +\frac{y^2}{2} f + g\lambda(t) \delta(y) f,
\label{eq:f}
\end{equation}
where $V(x)=g\delta(x)$ was used.
We see that the time dynamics in Eq.~(\ref{eq:f}) is driven by the parameter $g\lambda(t)$. Therefore, for very weak or strong interactions, where $g\lambda(t)$ scales out of the problem, Eq.~(\ref{eq:f}) becomes stationary. The same would also hold if $g$ was dependent on time such that $g=g_0/\lambda(t)$. 

It is worthwhile noting, that a scale transformation applied to an $N$-body problem yields an equation similar to Eq.~(\ref{eq:f}). Therefore, a gas in a time-dependent trap can be always mapped onto a gas with time-dependent interactions in a stationary trap, some properities of which were studied in Refs.~\cite{petrov2011, Ioannis2012}. 

{\it Weak interaction and/or weak perturbation of the trap}. It is straightforward to construct an iterative series for the function $f$ using Eq.~(\ref{eq:psi}).  Let us derive explicitly the first non-trivial term in this series in the limit of a small excitation parameter, i.e., $g\lambda(t)=g+g\Delta(t)$, where $\Delta(0)\equiv 0$ and $g|\Delta(t)|\ll 1, \forall t$.  This limit is reached if $g$ is small and we are far from parametric resonances, or if the trap is weakly perturbed.  

We look for a solution of the form
\begin{equation}
f(y,\tau)=e^{-i E \tau}f_0(y)+f_1(y,\tau),
\end{equation}
where $f_0(y)$ is an eigenstate of the stationary Hamiltonian (i.e., $H$ at $t<0$) of energy $E$. Note that $f(y,0)=f_0(y)$ and $f_1(y,0)=0$. The $f_0$ and $E$ are taken from the set of solutions $\{\varphi_i,E_i\}$ of a two-atom problem in a stationary trap \cite{busch1998}.   These solutions of even parity are written as
\begin{equation}
\varphi_{i,\mathrm{even}}(y)=N_{i} e^{-\frac{y^2}{2}}U(-\nu_i,\frac{1}{2},y^2), \; i=0,1,2,\dots ,
\label{eq:varphi_int}
\end{equation}
where $N_{i}$ is the normalization constant (see Appendix~\ref{sec:app}), $U$ is Tricomi's confluent hypergeometric function and $\nu_i=\frac{E_{i,\mathrm{even}}}{2}-\frac{1}{4}$ is one of the roots of the equation
\begin{equation}
\frac{\Gamma(-\nu_i+1/2)}{\Gamma(-\nu_i)}=-\frac{g}{2},
\label{eq:en_g}
\end{equation}
where $\Gamma(x)$ is the gamma function. To find the odd parity solutions, we note that the wave function in this case does not depend on the interaction. Indeed, the wave function vanishes at $x=0$, hence, the contribution of the interaction to the energy is exactly zero.  This means that the odd parity solutions for $g>0$ are the same as at $g=0$, therefore,  $\varphi_{k,\mathrm{odd}}=\phi_{2k+1}$, and the corresponding energy is $E_{k,\mathrm{odd}}=\frac{3}{2}+2k$, where $k=0,1,\dots\;$.  Notice that the parity of the wave function is conserved in a time-dependent oscillator. Therefore, if we expand the wave function in a time-independent basis $\{\varphi_i\}$ only the even (odd) states can be populated for an even (odd) initial state. The odd parity solutions vanish at zero, which means that the interaction does not affect the dynamics. Therefore,   the odd parity solutions are the same as in the non-interacting case,  and we do not consider them here. For convenience, we omit the subscript 'even' from now on. 

\begin{figure}
\centerline{\includegraphics[scale=0.85]{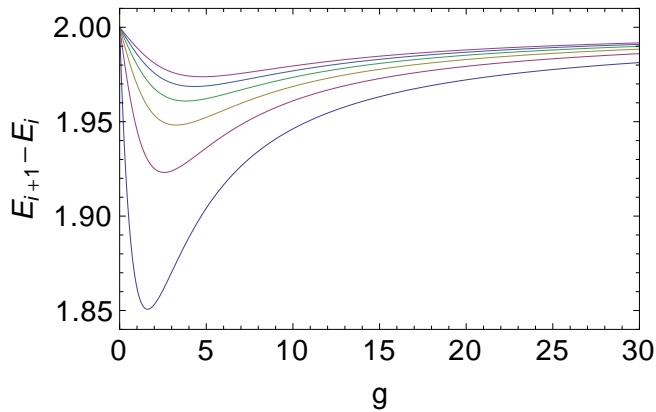}}
\caption{(Color Online). The energy differences between neighboring levels $E_{i+1}-E_{i}$
as functions of $g$ for $i=0,1,2,3,4,5$ (from bottom to top).}
\label{fig:energy_diff}
\end{figure}

The function $f_1$ satisfies the equation
\begin{equation}
i\frac{\partial f_1}{\partial \tau}=-\frac{1}{2}\frac{\partial^2 f_1}{\partial y^2}  +\frac{y^2}{2} f_1 + g(1 +\Delta) \delta(y) f_1 + g \Delta\delta(y)f_0,
\end{equation}
which should be supplemented with the initial condition $f_1(\tau=0)=0$. Having this condition in mind, we neglect the term $g\Delta(t)\delta(y) f_1$, which is
tiny in the beginning. In the resulting equation, the $g  \Delta\delta(y)f_0 $ term acts as the source, which leads to the following form of the $f_1$:
\begin{equation}
f_1(y,\tau)= - i f_0(0)\sum_{i} C_{i}(\tau)\varphi_{i}(y) \varphi_{i}(0),
\label{eq:f1}
\end{equation}
where $C_{i}(\tau)=g e^{-iE_{i}\tau}\int_0^\tau \mathrm{d}\tau'e^{i\tau' (E_{i}-E)}\Delta(t')$. This is the standard result for weak time-dependent perturbations (cf.~Ref.~\cite{landaubook}). Note that $C_i(\tau)$ is small unless $\Delta(t)$ has modes that oscillate in resonance with the energy differences in the system, i.e., $\Delta(t)\sim e^{i(E-E_i)\tau(t)}+\dots$. To obtain further insight, we plot these energy differences in Fig. \ref{fig:energy_diff}. We see that for $g=0$ and $1/g=0$ the energy differences are equal. This  ensures the appearance of parametric resonances for periodic driving even for $\alpha\to0$ (see Fig. \ref{fig:lambda2}).   For $0<g<\infty$, we see that $E_{i+1}-E_i$ is slightly smaller than two for every $i$ and that the deviation is largest for $i=0$ at $g\sim 1$. Furthermore, for higher energies the differences are less affected by the value of $g$.

Higher order terms (that appear when we consider the effect of the $g\Delta(t)\delta(y) f_1$ term) can be large only if $\Delta$ has oscillation modes with frequencies that contain an approximately integer number of characteristic energies. This is a necessary condition for the occurrence of quantum parametric resonances, for which a small perturbation can drastically change the shape of the wave function. 
However, for the non-resonant regime and small excitation parameter $\Delta$ we expect the wave function to be of the form\footnote{It is interesting to note that if $g\to\infty$ then $f_1\simeq1/g$ and accordingly $f\simeq f_0 e^{-iE\tau}+\frac{\phi}{g}$, cf.~Ref.~\cite{volosniev2015}. Therefore, one can use $1/g$ as a small parameter in this case.} $f_0 e^{-iE\tau} +f_1$, where $f_1$ is given by Eq.~(\ref{eq:f1}). In this case the dynamics is mainly driven by the time dependence of the trap. Furthermore, this dynamics can be inferred from the scale transformation, just as in the non-interacting case.

{\it Numerical Simulations}. For strong driving of the system we solve Eq. (\ref{eq:schr}) numerically. To do this, we expand the wave function $\Psi(x,t)$ in the eigenbasis, $\varphi_{i}(x)$, of the stationary Hamiltonian
\begin{equation}
\Psi(x,t)=\sum_{i} a_i (t)\varphi_{i}(x),
\label{eq:decomp}
\end{equation}
and solve a linear system of differential equations (see Appendix \ref{sec:app} for details) assuming that initially the system occupies the ground state, i.e., $a_0(0)=1$ and $a_{i\neq0}(0)=0$. 
 This choice is motivated by the recent realization of the two-atom system in the ground state of the trap~\cite{jochim2012}.

We start by studying the dynamics of the system for the exponential change of the frequency, Eq.~(\ref{eq:omega_exp}). Ostensibly, for this case $\mathrm{d} \epsilon/\mathrm{d}t=0$ if $t>T$. This implies that the amount of energy pumped into the system is finite, and consequently we do not expect the dimension of the basis, needed to obtain converged results, to be large.  We present our findings for the root-mean-square radius in Fig.~\ref{fig:re}, where the results for $g=1$ are plotted next to the non-interacting case. We see that the oscillations are slower in the interacting case. Indeed, the intrinsic time scales set by $1/(E_{i+1}-E_{i})$ are larger than in the non-interacting case, see Fig.~\ref{fig:energy_diff}. We carried out numerical simulations for different values of parameters, $\alpha$ and $T$ and observed the same trend. For values of $g$ that are not too close to one the difference between the non-interacting and interacting cases is minute. We also found that the behavior of other observables, such as $\epsilon(t)$, can be understood qualitatively from the $g=0$ case.

\begin{figure}
\centerline{\includegraphics[scale=0.85]{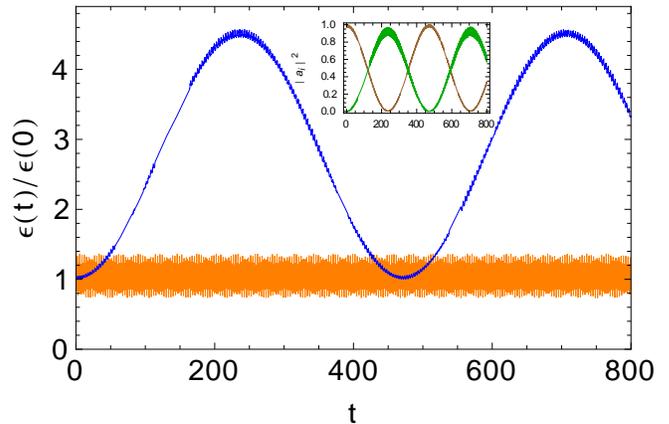}}
\caption{(Color Online). The energy, $\epsilon(t)/\epsilon(0)$, as a function of time for the periodic driving, $\omega^2 (t) = 1 + \alpha\sin(t/T )$, with $\alpha=0.5$, $T=1/3.80$ for $g=0$ (orange) lower curve and $g=2$ (blue) upper curve. The inset shows the population of states, $|a_i|^2$, for the $i=0$ and $i=2$ states in the case $g=2$. Note that $|a_0(0)|^2=1$ and $|a_2(0)|^2=0$. }
\label{fig:almostres}
\end{figure}

Let us now consider a periodic change of frequency, $\omega^2 (t) = 1 + \alpha\sin(t/T )$.  We start by considering the parameters from the white area in Fig.~\ref{fig:lambda2} which lead to a stable behavior in the non-interacting case. First of all, we note that in this region we do not expect parametric resonance to occur for any value of $g$. Indeed, at resonance the $\epsilon(t)$ becomes very large at long times, and because the scattering properties are set by the parameter $g/\sqrt{\epsilon(t)}$ the resonant system should behave effectively as non-interacting for $t\to\infty$ and $1/g\neq0$, see also Appendix~\ref{sec:appb}.  This can also be  understood from Fig.~\ref{fig:energy_diff}, where the energy differences are closer to the non-interacting value for larger values of $i$. Let us note that contrary to the case with exponential change of frequency, the behavior of the interacting system can be very different compared to the non-interacting counterpart. We illustrate this by considering $\alpha=0.5$, $T=1/3.80$ for $g=0$ and $g=2$. In the latter case, the energy scale given by $1/T$  
sets the ground state in resonance with the second excited state ($E_2-E_0\simeq3.80=1/T$), see Fig.~\ref{fig:energy_diff}. Furthermore, unlike the $g=0$ case, the transition element between these two levels does not vanish.  Therefore, we should observe an enhancement in the population in the second excited state. This is plotted in Fig.~\ref{fig:almostres}. Here we see that the system oscillates between the ground and the second excited states with only a marginal probability to be in other states. At the same time, the non-interacting system stays in the ground state and does not change its energy appreciably. Additionally, we note that outside the special regions where the driving is in resonance with the eigenstates of the interacting system, the behavior of the interacting and non-interacting systems is similar.

\begin{figure}
\centerline{\includegraphics[scale=0.45]{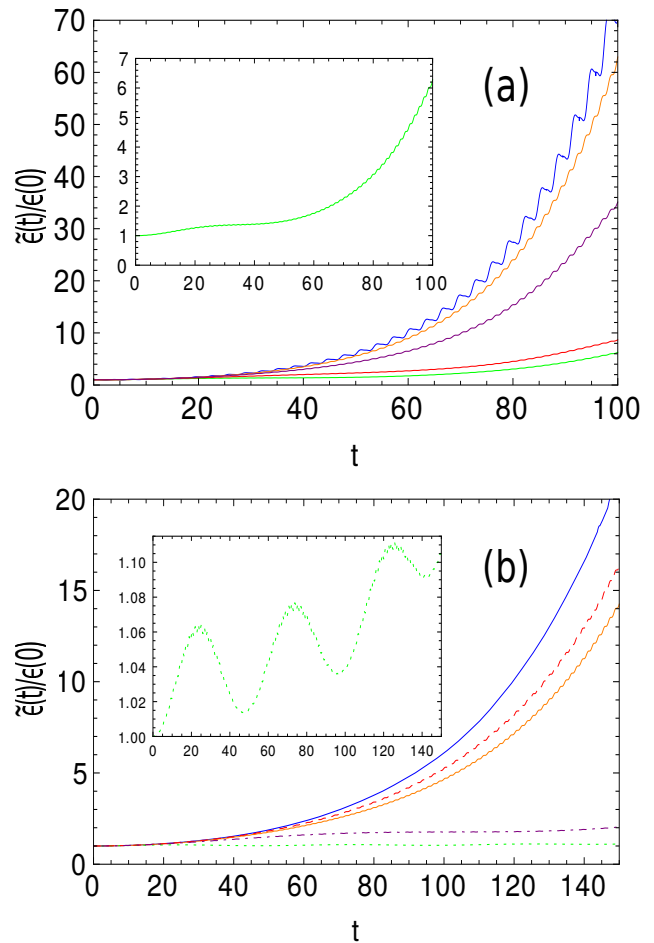}}
\caption{(Color Online). The energy $\tilde \epsilon(t)/\epsilon(0)$ as a function of time for periodic driving, $\omega^2 (t) = 1 + \alpha\sin(t/T )$. Panel~(a) corresponds to the case $\alpha=0.1$, $T=0.5$ where $g=0.1$ is shown with the (orange) solid curve, $g=1$ -- (green) dashed curve, $g=5$ -- (red) dotted curve, and $g=10$ -- (violet) dot-dashed curve. Panel~(b) depicts the case $\alpha=0.05$, $T=0.5$ where $g=0.1$ is shown with the (orange) solid curve, $g=1$ -- (green) dashed curve, $g=10$ -- (violet) dot-dashed curve and $g=30$ -- (red) dotted curve. The uppermost thin curves on both panels show the case $g=0$ ($1/g=0$). In the insets we zoom in for the case $g=1$.}
\label{fig:res_comb}
\end{figure}

Now we proceed to the interacting system with parameters taken from the instability region in Fig. \ref{fig:lambda2}. When the driving is much stronger than the energy effect induced by $V$ (see Fig.~\ref{fig:energy_diff}) our expectation is that the role of the interaction in the dynamics is marginal. As a consequence, we start our discussion by considering a moderate driving strength $\alpha=0.1$ for $T=0.5$. For this case we calculate the quantity $\tilde \epsilon(t)=\sum|a_i|^2 E_i$, see Fig.~\ref{fig:res_comb} (a), which equals to $\epsilon(t)$ if $t=\pi n T$ (cf.~Appendix \ref{sec:app}). Here we see that the interaction affects the resonance pattern significantly. Indeed, in the $g=0$ case we have $1/T=E_{i+1}-E_i$ which leads to a steep energy increase. For $g>0$ the resonance condition is fulfilled only for the highly excited states, which changes the dynamics as shown in Fig.~\ref{fig:res_comb} (a). This change is especially obvious for $g=1$, where the energy difference deviates notably from the $g=0$ ($1/g=0$) case (cf.~Fig.~\ref{fig:energy_diff}). If we make the driving even weaker and set $\alpha=0.05$ for the same $T=0.5$, the effect of interaction becomes even more pronounced, see Fig.~\ref{fig:res_comb} (b). Although, we see that for any interaction strength the energy increases,  the rate of this increase is strongly affected by the interaction. It reveals that the contribution of two-body interactions is important for the dynamics if $\alpha$ is small. This is in agreement with the experimental findings of Ref.~\cite{Wang2007}.

Our results suggest a way to use parametric excitations to cool the system even in an ideal harmonic trap.  Indeed, Fig.~\ref{fig:res_comb} implies that for weak driving particles with low collision energies should be heated less than the particles with high collision energies. Therefore, parametric excitations lead to a forced evaporative cooling\cite{Modugno2002}, which should be contrasted with the parametric heating or cooling due to the anharmonicity of the trap~\cite{thomas1997, Modugno2002}.  However, to make quantitative predictions about this process more detailed many-body investigations are required.   In general, the parametric resonance in realistic systems needs further studies. Indeed, as the energy becomes very large, the interaction cannot be described with the delta function and the trap with the oscillator potential anymore. Furthermore, the assumption that the system is one-dimensional can be accurate only for low energies, for which two dimensions are energetically frozen.
For high energies, three-dimensional calculations should be provided. We leave these investigations for future studies.

\begin{figure}
\centerline{\includegraphics[scale=0.85]{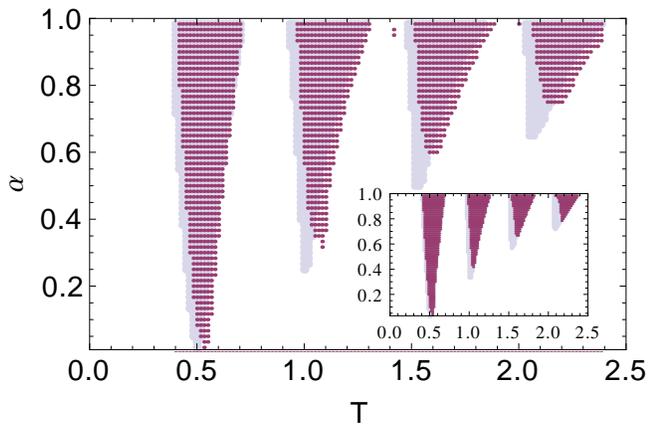}}
\caption{(Color Online). The parameter regions where $\tilde \epsilon (t_0)/\epsilon(0)-1>P\alpha$ for some $t_0<100$. In the main panel we show our results for $g=1, P=5$ using (red) points, the case $g=0, P=5$ ($1/g=0$) is depicted with shaded area. In the inset we present results for the same systems with $P=15$.}
\label{fig:st_new}
\end{figure}

Note, that the oscillator potential couples the initial state to every state in the basis, see Appendix~\ref{sec:app}. If the system is placed in a highly excited state then it should behave as non-interacting (indeed, in this case $g/\sqrt{\epsilon(t)}\to0$, because $\epsilon(0)$ is large). For that reason, we expect that the interacting system is resonant in the shaded region of Fig.~\ref{fig:lambda2}, where the non-interacting system is resonant, see also Appendix~\ref{sec:appb}. At the same time, Fig.~\ref{fig:res_comb} suggests that this result might be of limited use, because the signal can be affected strongly at experimentally relevant short times by the presence of interaction. 
Therefore, we leave the rigorous mathematical investigation of parametric resonances in the interacting systems for future studies. Instead, we look for the regions in the parameter space $(\alpha, T)$ for which within some holding time we can pump in energy into the system. More precisely, we look for the systems for which there is some $t_0<T_{max}$, such that  $\tilde \epsilon (t_0)/\epsilon(0)-1>P\alpha$. The choice of parameter $P$ is rather arbitrary and we take $P=5$ and $P=15$. 
Furthermore, we put $g=1$ and as a holding time we use $T_{max}=100$, which is much smaller than the lifetime of the current experimental set-ups (cf.~Ref.~\cite{jochim2012}), but much larger than the time scale given by the trap. We present our results in Fig.~\ref{fig:st_new} for $P=5$ in the main panel and $P=15$ in the inset. We notice that the interacting case, depicted with (red) points, is shifted towards larger values of $T$, which is due to smaller energy differences in this case, see Fig.~\ref{fig:energy_diff}. At the same time, when we increase the value of $P$, we do not find red points outside of the shaded region, see the inset. A closer investigation reveals that the red points outside of the shaded region for $P=5$ have dynamics similar to the one plotted in Fig.~\ref{fig:almostres}. This agrees with our expectation that in the interacting case resonance can occur only for the parameters from the instability region in  Fig. \ref{fig:lambda2}. Note that we expect similar results if the initial state is a low-lying excited state, e.g., $a_{i\neq1}(0)=0, a_{1}(0)=1$. If the initial state is a highly excited state then the effect of the interaction should be less pronounced. 

\section{Summary and Outlook}
\label{sec:concl}

We considered a two-body problem in a 1D harmonic trap with a time-dependent frequency. First, we studied the non-interacting case where the dynamics is determined by the classical counterpart. Next, we studied the effect of the zero-range interaction on the dynamics.  
We found that if the frequency is changed rapidly from one value to another the dynamics can be understood by proper rescaling of observables for any value of $g$.  At the same time, a periodic change of frequency can give rise to very different dynamics. In particular, the interaction can slow down the rate of the energy increase in the regions where the parametrical resonance phenomenon  can occur.

It would be interesting to consider extensions of this work to two- and three-dimensional geometries. There, in the non-interacting case the Hamiltonian is a sum of the 1D oscillators, and, therefore, the dynamics is determined by the equations presented above. 
Moreover, just as in the 1D case, the interacting system is solvable in a stationary trap for a zero-range pseudopotential~\cite{busch1998}. Therefore, the analysis of the one-, two- and three-dimensional systems can be provided along the same lines.
Another important extension is to consider attractive interactions. This case differs from the present one mainly due to the existence of a molecular state~\cite{busch1998}, which can have negative energy.

It might be worthwhile to study a classical analogue of the present problem, i.e., a harmonic oscillator with a short range potential in the middle. Of course, the problem will depend strongly on the form of the interaction and initial conditions. However, such an investigation could shed some additional light onto the quantum problem. 

The extension to the corresponding many-body problem would also be interesting. There, in agreement with our findings,  it is known that for a rapid change of frequency the dynamics of the wave function is mainly described by the evolution of the scaling parameters \cite{castin1996}. At the same time our study implies that for the resonant periodic driving the dynamics can be strongly affected by the interaction, especially if the driving is weak. This paves the way for a forced evaporative cooling method, which requires more detailed many-body investigations.  

\begin{acknowledgements}
We acknowledge fruitful discussions with Oleksandr Marchukov and Nikolaj Zinner.
Besides, we want to thank Wael Elkamhawy and Marcel Schmidt for their thoughtful comments on the manuscript.
This work was supported in part by the Helmholtz-Association under
contract HA216/EMMI.
\end{acknowledgements}

\appendix

\section{Details of the numerical simulations}
\label{sec:app}

To solve Eq.~(\ref{eq:schr}) we expand the wave function $\Psi(x,t)$ in the eigenbasis, $\{\varphi_{i}(x)\}$, from Eq.~(\ref{eq:varphi_int}) 
\begin{equation}
\Psi(x,t)=\sum_{i=0}^{N_B} a_i (t)\varphi_{i}(x),
\label{eq:decomp}
\end{equation}
 where $N_B$ is the number of basis states used.
The coefficients $a_{i}$ then satisfy the linear system
\begin{equation}
i\frac{\mathrm{d} a_{i}(t)}{\mathrm{d}t}=E_{i}a_{i}(t)+\frac{\omega^2(t)-1}{2}\sum_{i'}A_{i i'}a_{i'}(t),
\label{eq:line_sys_a}
\end{equation}
where $A_{i i'}=\int \mathrm{d}x \varphi_{i}(x) x^2 \varphi_{i'}(x) $.  For $g\to0$ ($1/g\to0$) the matrix $A$ is tridiagonal and can be easily calculated because $\varphi_i(x)$ are the standard harmonic oscillator wave functions. As we show below, the matrix $A$ can also be written analytically for $0<g<\infty$. 

Let us first write an analytic expression for $N_{i}\equiv1/\sqrt{\int\mathrm{d}x e^{-x^2} U(-\nu_i,\frac{1}{2},x^2)^2}$:
\begin{equation}
N_{i}=\sqrt{\frac{\Gamma(-\nu_i)\Gamma(-\nu_i+\frac{1}{2})}{\pi(\psi\left(-\nu_i+\frac{1}{2}\right)+\psi(-\nu_i))}},
\end{equation}
where $\psi(x)\equiv\Gamma'(x)/\Gamma(x)$ is the digamma function \cite{abram}. As follows from Eq. (\ref{eq:en_g}), this expression is well-defined for $0<g<\infty$. Next, we write the element $A_{ii}$ as
\begin{equation}
A_{ii}=E_{i}-\frac{g \pi N_{i}^2}{2 \Gamma\left(-\nu_i+\frac{1}{2}\right)^2}.
\end{equation}
Note that for large $i$ we have: $E_i\simeq 2i+\frac{1}{2}+\frac{g}{\pi\sqrt{i}}$ and $A_{ii}\simeq 2i+1+\frac{g}{2\pi\sqrt{i}}$.  We see that in this limit the effect of the delta-function interaction is determined by the parameter $g/\sqrt{i}$, which becomes negligible for large $i$.
Finally, we write an expression for $A_{ii'}$ ($i\neq i'$):
\begin{align}
A_{ii'}=\frac{2\pi N_{i} N_{i'}}{\left[(E_{i}-E_{i'})^2-4\right](E_{i}-E_{i'})} \nonumber \times \\ \left[\frac{E_{i}+3E_{i'}}{\Gamma(-\nu_i)\Gamma\left(-\nu_{i'}+\frac{1}{2}\right)}-\frac{E_{i'}+3E_{i}}{\Gamma(-\nu_{i'})\Gamma\left(-\nu_i+\frac{1}{2}\right)}\right].
\label{eq:aii}
\end{align}

We see that the matrix $A$ couples the initial state to every excited state.  Moreover, one can determine the couplings to the highly excited states by establishing the limit $\lim_{i\to\infty}(-1)^i A_{ii'} i^{9/4} = b$, where $b$ depends on $i'$ and~$g$.

\begin{figure}
\centerline{\includegraphics[scale=0.85]{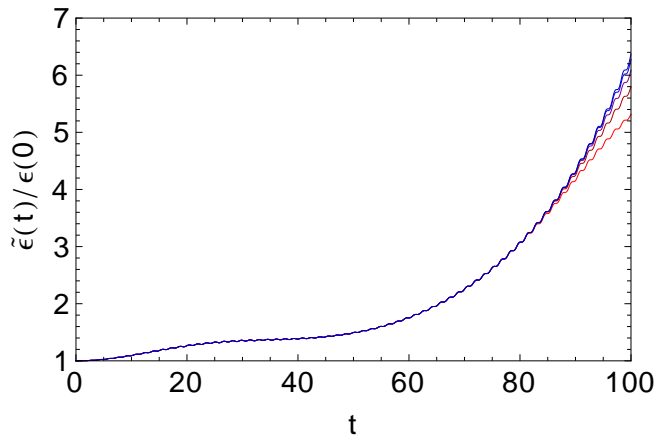}}
\caption{(Color Online). The energy $\tilde \epsilon(t)/\epsilon(0)$ as a function of time for periodic driving, $\omega^2 (t) = 1 + \alpha\sin(t/T )$, for $\alpha=0.1$, $T=0.5$ and $g=1$. The curves (from the bottom to top) correspond to basis sizes $N_B=100,125,150,175, 200$. }
\label{fig:convergence}
\end{figure}

The elements $A_{ii'}$ are very small for  $|i-i'|\gg 1$, thus, we truncate the matrix $A$ and solve linear system of equations (\ref{eq:line_sys_a}) using the standard routines from MATHEMATICA. To ensure accuracy of our analysis we perform calculations with different sizes of the matrix $A$.  As an example, we show such calculations, for a resonant case from Fig.~\ref{fig:res_comb}(a) corresponding to $\alpha = 0.1 , T = 0.5$ and $g=1$, in Fig~\ref{fig:convergence}. Here we plot the results for $N_B=100,125,150,175,200$ from the bottom to top. We see that for $t\lesssim80$ the curves coincide, and therefore any of the considered $N_B$ can be used. Note that the basis size needed to obtain converged results for a given time interval increases fast with the size of this interval.  For example, at $t\lesssim90$ curves for $N_B=175$ and $N_B=200$ coincide, whereas the relative difference between $N_B=175$ and $N_B=200$ at $t=100$ is already around one percent. 

The increase of $N_B$ is due to the exponential growth of the energy at resonance. Therefore, to study the behavior at large times a different numerical approach should be used. For instance, one can try to combine analytical results for high energy with the numerical results for low-excited states. This approach is left for future studies. Instead, we work with $N_B=175$ which ensures a reasonable calculation time\footnote{In our implementation, $N_B=175$ ($N_B=200$) assumes around 30 (45) minutes one kernel calculations per every curve in Fig.~\ref{fig:res_comb}} and good accuracy. To be on the safe side, we checked that $N_B=175$ reproduces the analytical non-interacting case for the parameters in Fig.~\ref{fig:res_comb} and for some random points from Fig.~\ref{fig:st_new}. It is worthwhile noting that the resonant cases are demanding from the numerical point of view because the system populates high energy states. For a non-resonant system the higher excited states are not populated and accurate results can be obtained with less effort.

For convenience, we provide expressions for $r(t)$ and $\epsilon(t)$ used in the main text
\begin{align}
&r^2(t)=\sum_{i,j}A_{ij}a_i^*a_j, \;\; \\
&\epsilon(t)=\tilde \epsilon (t) + \frac{\omega^2(t)-1}{2}r^2(t), 
\end{align}
where $\tilde \epsilon(t)=\sum|a_i|^2 E_i$. Note, that for periodic driving  we only consider and draw $\tilde \epsilon(t)$, because the last term in $\epsilon(t)$ is an oscillatory function.

\section{On the parametric resonance region for the interacting system}
\label{sec:appb}

In the main text we presented simple physical arguments that lead to the conclusion that the occurrence of parametric resonances does not depend on $g$. This conclusion rests on the assumption that if the system initially populates a highly excited state of the non-interacting stationary Hamiltonian, then the time evolution is mainly driven by the non-interacting piece of the Hamiltonian. In this appendix, we present some arguments to justify this statement. For this we consider Eq.~(\ref{eq:f}) and look for a solution in the standard iterative form which can be obtained from Eq.~(\ref{eq:psi}):  $f(y,\tau)=F_0(y,\tau)+F_1(y,\tau)+...$, where  the initial condition is $f(y,\tau)=F_0(y,0)$. Note, that by construction the function $F_0(y,\tau)$ solves the non-interacting Hamiltonian with the same initial condition, i.e., $F_0(y,\tau)=e^{-iE\tau}F_0(y,0)$. 
In this appendix we argue that the effect of $F_1$ is small.  However, it should be noted that a detailed investigation of the regions where parametric resonances appear is out of scope of the present paper and left for future studies. 

Repeating the steps from the main text we derive 
\begin{equation}
F_1(y,\tau)=-i g F_0(0,0)\sum_i \phi_i(y)\phi_i(0) \tilde C_i(\tau),
\end{equation}
where $\tilde C_i(\tau)=e^{-iE_i\tau}\int_0^\tau \mathrm{d}\tau' e^{i\tau'(E_i-E)} \lambda(t')\mathrm{d}\tau'$.
At first glance it might seem that $\tilde C_i(\tau)$ can diverge for $\tau\to\infty$, because $\lambda(t)$ becomes  exponentially large at large times. However, Fig.~\ref{fig:tau} suggests that the effect of $\lambda$ is suppressed in the $\tau-$domain. Let us investigate this effect in more detail. For this we consider the integral
\begin{equation}
I(t_1,t_2)=\int_{\tau(t_1)}^{\tau(t_2)} \lambda(t')\mathrm{d}\tau'= \int_{t_1}^{t_2} \frac{\mathrm{d}t'}{\lambda(t')}.
\end{equation}
Note that $|\tilde C_i(\tau)|\leq I(0,t)$. The main contribution to $I(0,t)$ comes from the region where $\lambda$ is small. 

For large energies, $\epsilon$, the time that a particle spends close to the origin becomes shorter. To better understand the behavior of $I(t_1,t_2)$ in this case, we approximate $\lambda(t)$ with $\tilde \lambda(t) = \mathcal{A} \sqrt{ \sin^2(t)+\cos^2(t)/\mathcal{A}^4}$. The function $\tilde \lambda(t)$ is the solution to Eq.~(\ref{eq:lambda}) for $\omega^2(t)=1$ that satisfies the initial conditions $\tilde \lambda(0)=1/\mathcal{A}$ and $\dot {\tilde \lambda}(0)=0$. Note, that $\mathcal{A}$ sets the energy scale of the system.  Now let us consider the integral
\begin{equation}
\tilde I (n)\equiv\int_{n\pi}^{n\pi+\pi/2}\frac{\mathrm{d}t'}{\tilde \lambda(t')}=\mathcal{A} K(1-\mathcal{A}^4),
\end{equation}
where $K(m)=\int_0^{\pi/2}\mathrm{d}\theta/\sqrt{1-m\sin^2\theta}$ is the complete elliptic integral of the first kind. Therefore, for large $\mathcal{A}$ we have $\tilde I(n)\sim \frac{\mathrm{ln} (\mathcal{A})}{\mathcal{A}}$. This should also quantify $I(n\pi,n\pi+\pi/2)$ for large $n$, because the behavior close to the origin is not affected strongly by the time-dependent oscillator. This observation implies that $I(0,t)$ is a bounded function. It is also interesting to calculate the integral $\int_{n\pi}^{n\pi + \pi/2} \mathrm{d}t' /\tilde \lambda(t')^2=\pi/2$. This result is in a good agreement with our observation that $\tau(t+n\pi)-\tau(t)\sim n\pi$.

Next we would like to show that the overlap $O=\langle F_1| F_1 \rangle$ is small if $(g F_0(0,0))^2\to 0$.  It is obvious that 
\begin{equation}
O=(g F_0(0))^2 \sum_i |\phi_i(0)|^2 |\tilde C_i(\tau)|^2.
\end{equation}
Smallness of $O$ can be deduced by noticing that the sum converges. Indeed, $I(n\pi,n\pi+\pi/2)$ is exponentially small for large $n$, thus, to obtain $I(0,t)$ we need to integrate only up to some $t_0$. Therefore, the integral in $\tilde C_i(t)$ is highly suppressed for large $i$. In a similar manner one can show that the higher order terms, i.e., $F_2, F_3\dots$, are small if $(g F_0(0,0))^2\to 0$.

\bibliographystyle{andp2012}
\bibliography{bib}

 \end{document}